\documentclass{INTERSPEECH2023}

\interspeechcameraready 
\usepackage{hyperref}

\title{Large-Scale Automatic Audiobook Creation}
\name{Brendan Walsh*$^1$, Mark Hamilton*$^{1,2}$, Greg Newby$^3$, Xi Wang$^1$, Serena Ruan$^1$, Sheng Zhao$^1$, \\ Lei He$^1$, Shaofei Zhang$^1$, Eric Dettinger$^1$, William T. Freeman$^{2,4}$, Markus Weimer$^1$}

\address{
  $^1$Microsoft, $^2$MIT, $^3$Project Gutenberg, $^4$Google, *Equal Contribution
  }
\email{}

\begin{document}

\maketitle

\begin{abstract}

An audiobook can dramatically improve a work of literature's accessibility and improve reader engagement. However, audiobooks can take hundreds of hours of human effort to create, edit, and publish. In this work, we present a system that can automatically generate high-quality audiobooks from online e-books. In particular, we leverage recent advances in neural text-to-speech to create and release thousands of human-quality, open-license audiobooks from the Project Gutenberg e-book collection. Our method can identify the proper subset of e-book content to read for a wide collection of diversely structured books and can operate on hundreds of books in parallel. Our system allows users to customize an audiobook's speaking speed and style, emotional intonation, and can even match a desired voice using a small amount of sample audio. This work contributed over five thousand open-license audiobooks and an interactive demo that allows users to quickly create their own customized audiobooks. To listen to the audiobook collection visit \url{https://aka.ms/audiobook}.

\end{abstract}

\section{Introduction}
Audiobooks have become a popular way to consume literature, news, and other publications. Audiobooks not only allow existing readers to be able to enjoy content on the go, but can help make content accessible to communities such as children, the visually impaired, and new language learners. Traditional methods of audiobook production, such as professional human narration or volunteer-driven projects like LibriVox, are time-consuming, expensive, and can vary in recording quality. These factors make it difficult to keep up with an ever-increasing rate of book publication. In contrast, automatic audiobook creation is orders of magnitude faster, cheaper, and more consistent but has historically suffered from the robotic nature of text-to-speech systems and the challenge of deciding what text should not be read aloud (e.g. tables of contents, page numbers, figures, and footnotes). 

We present a system that overcomes both of the aforementioned challenges by generating high-quality audiobooks from heterogeneous collections of online e-books. In particular, our system combines recent advances in neural text-to-speech, emotive reading, scalable computing, and automatic detection of relevant text to create thousands of reasonable-sounding audiobooks. We contribute over five thousand audiobooks totaling approximately thirty-five thousand hours of speech to the open source. We also contribute a demonstration app that allows conference attendees to create a custom audiobook, read aloud in their own voice, from any book from the collection using only a few seconds of example sound. 

\section{Related Work}

LibriVox is a well-known project that creates open-license audiobooks using human volunteers. Although it has made significant contributions to the accessibility of audiobooks, the quality of the produced audiobooks can be inconsistent due to the varying skills and recording environments of the volunteers. Furthermore, the scalability of the project is limited by the availability of volunteers and the time it takes to record and edit a single audiobook. Private platforms such as Audible create high-quality audiobooks but do not release their works openly and charge users for their audiobooks. Project Gutenberg hosts a broad collection of free e-books and a few audiobooks. Their existing audiobooks feature a robotic text-to-speech voice which limits listen-ability.  

Text-to-speech is a well-studied problem and recent deep learning methods such as WaveNet \cite{oord2016wavenet}, Tacotron \cite{Wang17-Tacotron}, and Fastspeech \cite{Ren19-Fastspeech} have shown considerable progress towards generating speech that rivals human quality and naturalness. In contrast, the problem of selecting which text to read from an e-book has received considerably less attention. Nevertheless, recent work by \cite{bodapati2018machine} has explored whether it's possible to predict the ``start reading location'' using LSTM-based models but does not tackle the cleaning of other irrelevant text throughout the body of an e-book.


\section{Methods}

This work introduces a scalable system capable of converting HTML-based e-books to high-quality audiobooks. Our pipeline is built using SynapseML\cite{pmlr-v82-hamilton18a}, a scalable machine learning framework that enables distributed orchestration of the entire audiobook creation process.

\subsection{Parsing e-Book HTML}

\begin{figure}[t]
    \centering
    \includegraphics[scale=0.23]{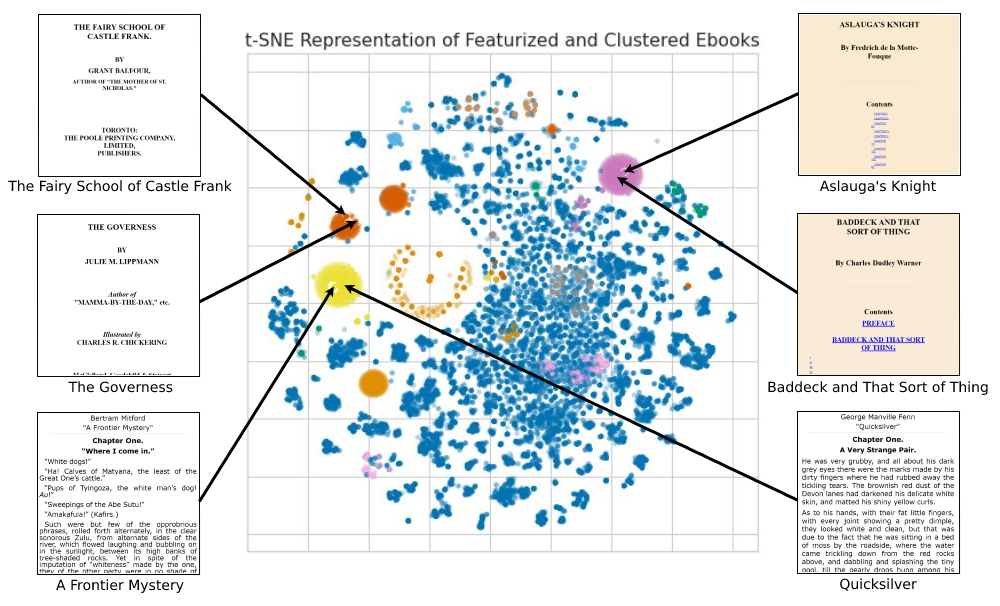}
    \vspace{-.2in}
    \caption{t-SNE Representation of Clustered Ebooks. Colored areas represent uniformly formatted clusters of books.}
    \vspace{-.25in}
    \label{fig:tsne}
\end{figure}

Our pipeline begins with thousands of free e-books provided by Project Gutenberg. These e-books are provided in several different formats, and our work focuses on their HTML format which is most amenable to automated parsing. Parsing this extremely heterogeneous and diverse collection of e-books was the most significant challenge we encountered. Project Gutenberg does not standardize the contents of its HTML files and its e-books contain a significant amount of text that would not be relevant for audio readers including pre-ambles, tables of contents, tables, illustrations, in-text page numbers, footnotes, transcriber notes, and other strange artifacts. 

To create a high-quality subset of e-books we first featurize each e-book's HTML Document Object Model (DOM) tree using a combination of automated (the TF-IDF statistic on HTML Components) and hand-crafted HTML features. This allowed us to cluster and visualize the entire collection of Project Gutenberg HTML files and allowed us to find several large groups of commonly structured files. We used these clusters of HTML files to build a rule-based HTML normalizer that converted the largest classes of e-books into a standard representation that could be automatically parsed. This analysis allowed us to create a system that could quickly and deterministically parse a large number of books. Most importantly it also allowed us to restrict attention to a subset of files that would generate high-quality recordings when read. 
Figure \ref{fig:tsne} shows the results of this clustering procedure, demonstrating that several clusters of similarly structured e-books naturally emerge in the Project Gutenberg collection. 
Once parsed we can extract a stream of plain text to feed to text-to-speech algorithms.



\subsection{Generating High Quality Speech}


Different audiobooks require different reading styles. Nonfiction works benefit from a clear and neutral voice while fictional works with dialogue can benefit from an emotive reading and some ``acting''. For the majority of the books, we use a clear and neutral neural text-to-speech voice, However, in our live demonstration we will present users with the ability to customize the voice, speed, pitch, and intonation of the text. 

To clone a user's voice we utilize zero-shot text-to-speech methods \cite{Wu22-Ada} 
to efficiently transfer the voice characteristics from limited enrolled recordings. This allows a user to quickly create an audiobook in their own voice using a small amount of recorded audio. 

To create an emotive reading of the text, we use an automatic speaker and emotion inference system to dynamically change the reading voice and tone based on context. This makes passages with multiple characters and emotional dialogue more life-like and engaging. To this end, we first segment the text into narration and dialogue and identify the speaker for each dialogue section. We then predict the emotion of each dialogue using \cite{Wu22-selfcontext} in a self-supervised manner. Finally, we assign separate voices and emotions to the narrator and the character dialogues using the multi-style and contextual-based neural text-to-speech model proposed in \cite{Guo21-conv}.

\section{The Project Gutenberg Open Audiobook Collection}

We introduce the Project Gutenberg Open Audiobook Collection: over five thousand high-quality audiobooks generated from the Project Gutenberg collection and available for free download and open use. We host these files as a single zip file for the research community as well as on the major podcast and audio file hosting platforms for use by the broader community. This collection offers over thirty-five thousand hours of content including classic literature, non-fiction, plays, and biographical works narrated in a clear and consistent voice. We hope this contribution can provide value to both the research community, and the broader community of audiobook listeners.

\section{Demonstration}

We plan to host a live demonstration application that allows conference attendees to create their own custom audiobooks using our system. Users will first start by selecting a book from the 5,000 titles in our collection using a simple search interface. They can then select what voice they would like to use for the recording from a large collection of existing neutral and emotion-aware voices or even their own voice. If a user wants to create a custom audiobook using their own voice, they will be asked to speak a few sentences to quickly train a custom voice profile. Users will be able to listen to a preview of their audiobook in real time and add an optional custom dedication before submitting a larger job that reads the entire book. Once the pipeline finishes we will email the user a link to download their custom-made audiobook.






\section{Conclusions}

In this work, we present a novel pipeline to automate the creation of high-quality audiobooks from heterogeneous e-books. Our system uses new advances in neural text-to-speech, emotion recognition, custom voice cloning, and distributed computing to create engaging and lifelike audiobooks. We apply this system to donate over five thousand audiobooks to the open-source community and aim to demonstrate this system by allowing conference attendees to create custom audiobooks. We believe that this work has the potential to greatly improve the accessibility and availability of audiobooks.



\bibliographystyle{IEEEtran}
\bibliography{mybib_abbrev}

\end{document}